\documentclass[aps,reprint,showpacs,superscriptaddress,floatfix]{revtex4-1}
\usepackage{comment}
\usepackage{graphics,amssymb,amsmath,epsfig,color,textgreek}
\usepackage{graphicx}
\usepackage[dvipsnames]{xcolor}
\usepackage{dcolumn}
\usepackage{bm}
\usepackage[colorlinks=true,citecolor=cyan]{hyperref}
\hypersetup{colorlinks=true,citecolor=cyan,linkcolor=red,urlcolor=magenta}
\usepackage{braket}
\usepackage[normalem]{ulem}
\usepackage{cancel}
\usepackage{diagbox}
\usepackage{lipsum}
\usepackage{cleveref}
\usepackage[colorinlistoftodos]{todonotes}

\crefformat{figure}{Fig.~#2#1#3}
\crefformat{equation}{Eq.~#2#1#3}
\crefformat{appendix}{App.~#2#1#3}

\usepackage[dvipsnames]{xcolor}
\definecolor{fg}{rgb}{1, 0, 0}

\definecolor{fb}{rgb}{0.7, 0.15, 0.6}
\newcommand{\fb}[1]{\textcolor{fb}{#1}}

\definecolor{eh}{rgb}{0.95, 0.15, 0.35}
\newcommand{\eh}[1]{\textcolor{eh}{#1}}

\definecolor{ng}{rgb}{0.7, 0.7, 0.2}

\definecolor{hl}{rgb}{0.6, 0.3, 0.9}

\newcommand{\LK}[1]{\textcolor{blue}{#1}}

\newcommand{\ham}{\mathcal{H}}

\newcommand{\II}{\mathcal{I}_2}
\newcommand{\CC}{\mathcal{C}_2}
\newcommand{\UU}{\mathcal{U}_2}

\newcommand{\bi}{\begin{itemize}}

\newcommand{\imag}{i}
\newcommand{\CARPES}{C-ARPES }

\newcommand{\kone}{{{k}_1}}
\newcommand{\ktwo}{{{k}_2}}

\newcommand{\qq}{{q}}
\newcommand{\kk}{{k}}
\newcommand{\pp}{{p}}

\begin{document}

\title{Observing two-electron interactions with correlation-ARPES}

\author{A.~F.~Kemper}
\email{akemper@ncsu.edu}
\affiliation{Department of Physics, North Carolina State University, Raleigh, North Carolina 27695, USA}

\author{Francesco Goto}
\affiliation{Centre \'{E}nergie Mat\'{e}riaux T\'{e}l\'{e}communications, Institut National de la Recherche Scientifique, Varennes, Qu\'{e}bec J3X 1S2, Canada}

\author{Heba.~A.~Labib}
\affiliation{Department of Physics, North Carolina State University, Raleigh, North Carolina 27695, USA}

\author{Nicolas Gauthier}
\affiliation{Centre \'{E}nergie Mat\'{e}riaux T\'{e}l\'{e}communications, Institut National de la Recherche Scientifique, Varennes, Qu\'{e}bec J3X 1S2, Canada}

\author{E.\,H.\,da Silva Neto}
\affiliation{Department of Physics, Yale University, New Haven, Connecticut 06511, USA}
\affiliation{Energy Sciences Institute, Yale University, West Haven, Connecticut 06516, USA}
\affiliation{Department of Applied Physics, Yale University, New Haven, Connecticut 06516, USA}

\author{F.\,Boschini}
\email{fabio.boschini@inrs.ca}
\affiliation{Centre \'{E}nergie Mat\'{e}riaux T\'{e}l\'{e}communications, Institut National de la Recherche Scientifique, Varennes, Qu\'{e}bec J3X 1S2, Canada}

\date{\today{}}

\begin{abstract}

Identifying and studying the underlying two-electron interactions that give rise to emergent phenomena 
is a key step in developing a holistic understanding of quantum materials.  This step is hindered by the lack of
an experiment that can directly interrogate the interactions and the specific quasiparticles involved
in the interaction simultaneously.  We introduce correlation-ARPES (C-ARPES) as a new experimental
method that overcomes this difficulty and directly measures the interactions between specific, chosen quasiparticles
by measuring the correlations between two electrons photoemitted
from the same pulse (but not the same photon).  We illustrate how this technique can extract the underlying
interactions, and demonstrate this with an example of phonon-mediated electron-electron interactions, such as
those that give rise to charge density waves through the Kohn anomaly mechanism.  

\end{abstract}

\maketitle

\section{Introduction}
Gaining insight into the many-body interactions between electrons
is a central theme in the quest to comprehend quantum materials. 
Many spectroscopic approaches have been developed and used to characterize and understand the many-body interactions that lead to emergent phenomena. 
Yet, while condensed matter scientists may dream of one day accessing all possible many-electron correlations in a solid, our understanding of interactions between merely two electrons remains notably limited, as we either observe only the echoes of many-body interactions through their effects on single-particle properties or only catch a glimpse of the interactions in an averaged sense, without being able to resolve details.

\begin{figure*}[htpb]
    \centering
    \includegraphics[clip=true,trim=45 20 45 30,
        width=0.99\textwidth]{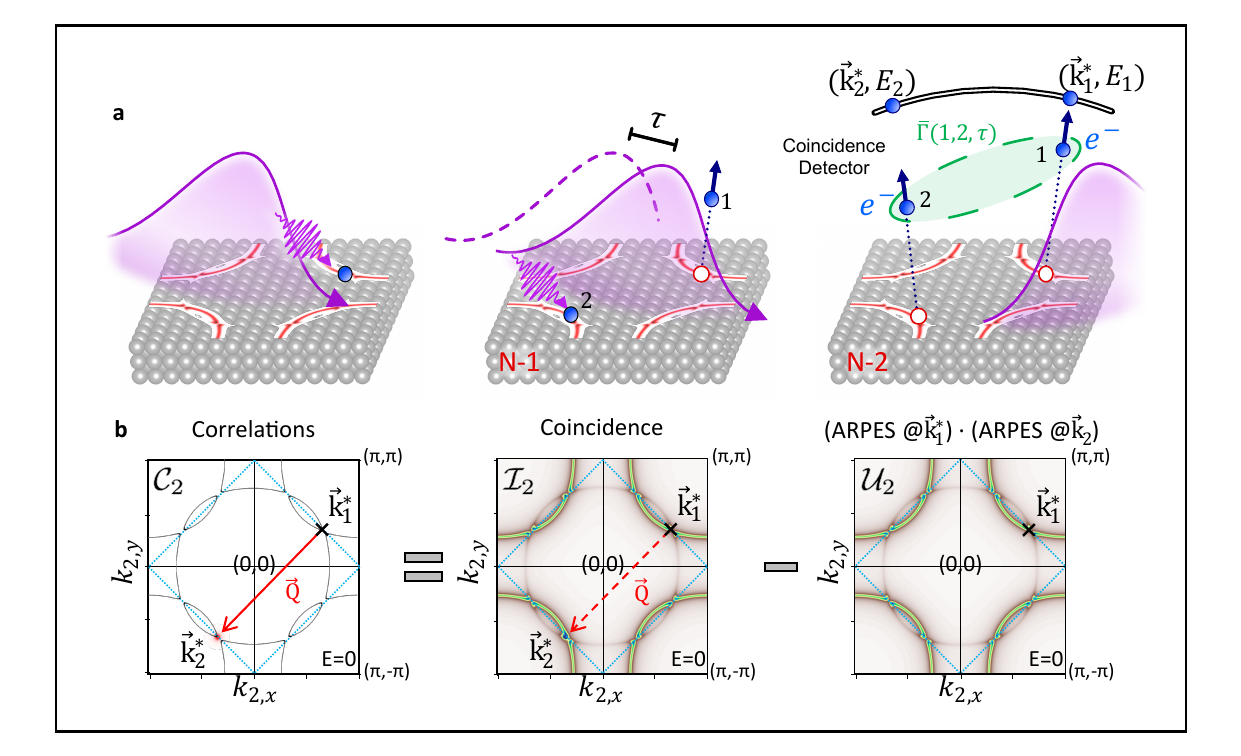}
    \caption{a) Schematic illustration of the correlation-ARPES (C-ARPES) process. Two electrons are photoemitted between the same pulse, and interactions between them
    are imprinted on the correlation signal $\CC$.
    b) Correlations ($\CC$) between the two photoemitted electrons $\Gamma_{k_1,k_2}$ are obtained by comparing the coindicence ($\II$)
    signal to the one-electron photoemission process ($\UU$).}
    \label{fig:schematic}
\end{figure*}

Two leading techniques used to study quantum materials today are
angle-resolved photoemission spectroscopy (ARPES), a photon-in/electron-out technique that provides direct insight into quasiparticle strength and dispersion in solids \cite{sobota2021angle,boschini2024time}, and resonant inelastic X-ray scattering (RIXS), an all-optical technique that reveals mechanisms of
momentum and energy transfer in the solid\cite{devereaux2007inelastic,ament2011resonant, Mitrano_Dean_PRX_RIXS_review_2014, NatRev_RIXS_review,daSilvaNeto_Frano_Boschini_Frontiers_RIXS_review}. These two techniques illustrate the
blind spot that exists in our understanding of interactions in a material due to the limitations of the experiments. ARPES measures the properties of electrons in the material as a function of their momentum and energy ($\kk, E$). 
Moreover, the single-particle spectral function probed by ARPES encodes information about the interactions between electrons or with a boson of some sort (emergent or not), characterized by momentum and energy $(q, \Omega)$. However, this information is not directly observed in the ARPES experiment; instead, it must be inferred from the dispersion and linewidth, interpreted through a hypothesized underlying model for the interaction boson \cite{lanzara2001evidence,johnson2001doping,lee2007aspects,ZONNO2021,iwasawa2013true,byczuk2007kinks,Kinks,boschini2018collapse,na2019direct}
On the other hand, RIXS can directly measure the bosonic momentum and energy ($q,\Omega$), but the information is averaged over the various ($\kk, E$) electronic quasiparticles involved in the cross-section that generates the bosonic spectral feature\cite{devereaux2016directly, Dashwood_PRX_2021, zinouyeva2025momentum}, with a model once again being hypothesized and its parameters adjusted to match the scattering experiment.
In both cases, the models are limited in their capacity to account for interaction effects, as they cannot directly capture both quasiparticle ($\kk, \omega$) and bosonic ($q,\Omega$) degrees of freedom simultaneously. 
There is thus a clear opportunity for an innovative method capable of directly probing correlations between two quasiparticles of interest --- 
$(\kone,E_1)$ and $(\ktwo,E_2)$, and their relative difference
$(q=\kk_2-\kk_1,\Omega = E_2-E_1)$ ---
thus unraveling the electron-electron interactions and overcoming the limitations
of ARPES and RIXS.

The development of new experimental methods to directly access the two-electron correlation function $\Gamma_{k_1,k_2}$ over the entire space of two-electron
states
$(k_1,E_1,q=k_2-k_1,\Omega=E_2-E_1)$ 
would have a tremendous impact on our understanding of why and under what conditions quantum phases of matter emerge. [Note that in $\Gamma_{k_1,k_2}$ we are indexing the interaction just by momenta for brevity, dropping the implicit time or energy variable.]
An approach to extract both quasiparticle and bosonic information within a single experiment is to build on our established understanding of the photoemission process.
For example, the extension of ARPES into the time-domain (time-resolved ARPES \cite{boschini2024time}) has recently been  proposed as a potential approach to extract $\Gamma_{k_1,k_2}$. By selectively optically exciting electrons in the surrounding of the $(\kone,E_1)$ electronic state and then tracking their ultrafast direct decay into $(\ktwo,E_2)$, \textcite{na2019direct} have extracted the mode-projected electron-phonon matrix element in graphite. However, although promising, this approach relies heavily on the availability of selective optical transitions, hence limiting its broad applicability.

Another ARPES based approach is to measure
the one-photon-in/two-electrons-out process (2e-ARPES), which has been performed in the pioneering work of the research groups of Kirschner, Schumann, and Widdra, demonstrating their potential in uncovering electron correlations in metals\cite{schumann2006mapping,schumann2007correlation,schumann2009sensing,munoz2009electron,huth2014electron,kostanovskiy2015core,schumann2016electron,trutzschler2017band,chiang2020laser}.
When looking at low-energy scales ($<100$ meV), 2e-ARPES could also enable direct access to electron interactions responsible for gluing electrons into Cooper pairs\cite{trutzschler2017band,mahmood2022distinguishing}.
However, 
the 2e-ARPES cross-section is expected to be much lower compared to a standard ARPES experiment, thereby introducing challenges in its experimental implementation\cite{chiang2020laser,leitner2021coesca,zwettler2024extreme}.

\begin{figure*}[htpb]
    \includegraphics[width=0.99\textwidth]{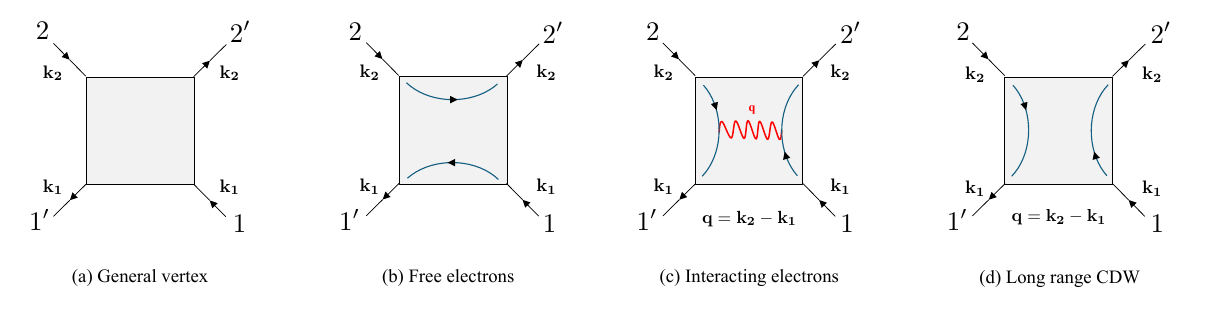}
    \caption{Feynman diagrams $G_2$ illustrating the 4-point vertex for the scattering process, where the primed (unprimed) times correspond to annihilation (creation) operators and
    have been drawn as outgoing (incoming) lines.}
    \label{fig:pendiagram1}
\end{figure*}

In this work, we outline how a new technique that we term correlation ARPES (C-ARPES) that can enables a comprehensive mapping of dynamic interactions in quantum materials by directly interrogating the electronic states involved. This method was first introduced by Stahl et
al.\cite{stahl2019noise} as a pump-probe coincidence ARPES experiment, who investigated noise-correlations in
ARPES and demonstrated the use of this concept in investigating
light-quenched superconductivity.
As we will show, \CARPES is much more general;
\CARPES probes the correlation between two electrons in the solid via the simultaneous detection of two electrons emitted by two photons belonging to the same ultrashort ultraviolet (UV) pulse, as sketched in~\cref{fig:schematic}a.
The \CARPES signal is proportional to the two-particle correlation function between the two quasiparticles and can be visualized in momentum space, \cref{fig:schematic}b. We show that if an electron with fixed momentum $k_1^*$ is correlated to another electron having momentum $k_2^*$, the  \CARPES signal $\CC \propto \Gamma_{k_1^*,k_2^*}$ can be obtained directly by evaluating the difference between the total number of two-electron coincidence events, $\II$, and the uncorrelated part, $\UU$ (which is proportional to the product between the ARPES intensities of two non-coincident single electrons).

In the following, we discuss the general theoretical framework of C-ARPES, and offer concrete examples.
We consider phonon-mediated electron scattering, and illustrate how this type
of experiment can reveal the Kohn anomaly that eventually leads to charge
density wave formation. We also briefly discuss the use of \CARPES to
study superconductors and charge density waves in the long-range ordered phase.

\section{Results}

In this section, we will outline the formalism that
underpins the study of C-ARPES,  making use of
the notation introduced by Stahl et al.\cite{stahl2019noise}.
 \CARPES measures the correlation for two coincident free electrons
with momenta $\pp$ and $\pp'$, energies $E_\pp$ and $E_\pp'$, and spins $\sigma$ and $\sigma'$:
\begin{align}
    \II = \langle \mathbf{S}^\dagger n^f_{\pp\sigma} n^f_{\pp'\sigma'} \mathbf{S} \rangle_0,
    \label{eq:I2_definition}
\end{align}
where $\mathbf{S}$ is the standard time-ordered $\mathbf{S}$-matrix\cite{Mahan,bruus}
\begin{align}
    \mathbf{S} = T_t e^{-i \int_{-\infty}^\infty d\bar t \ham(\bar t)}
\end{align}
and we use the time-loop $\mathbf{S}$-matrix formalism.
For comparison, standard single-electron ARPES signal $\mathcal{I}_1$ has
a single number operator $n^f_{\pp\sigma}$ in \cref{eq:I2_definition}.
Here, we will treat the $\mathbf{S}$-matrix in the interaction
representation with respect to two terms:
\begin{align}
    \mathcal{H}(t) = 
    H_{\text{p.e.}}(t)
    + H_{\mathrm{int}}
\end{align}

The first term is responsible for
electron emission in the sudden approximation\cite{damascelli2003angle,sobota2021angle,boschini2024time},
\begin{align}
    H_{\text{p.e.}} =
    &\sum_{\pp,\kk,\sigma,\sigma'} S(t) e^{-i\Omega t} M^{\sigma,\sigma'}_{k,p} c^\dagger_{\kk\sigma} f_{\pp\sigma'} + h.c.
\end{align}
Here, $\Omega$ is the photon energy, $c_{\kk\sigma}$ and $c^\dagger_{\kk\sigma}$ are the operators for the electrons in the solid, $f_{\pp\sigma'}$ and $f^\dagger_{\pp\sigma'}$ for the free electrons, and $S(t)$ is the probe pulse temporal envelope. $M^{\sigma,\sigma'}_{k,p}$ is the matrix element,
which for simplicity we will set to $M \delta_{k,p} \delta_{\sigma,\sigma'}$
(note that we will label the quasiparticles by their crystal momentum $\kk$).
More complex matrix elements can
be readily included, but occlude the notation.

The remaining interaction term 
depends on the particular interaction that is being
considered, \emph{e.g.} electron-electron or electron-phonon
scattering, which we will detail below.

\subsection{\CARPES of free electrons}

We first consider the simplest case of free electrons, with $H_{\text{int}}=0$.
A straightforward
perturbation theory in $H_{\mathrm{p.e.}}$, considering the initial free electron state to be empty, yields the conclusion
that the electron operators appear in a
particular order for this measurement cross-section \cite{stahl2019noise}. The second order perturbation theory in
$H_{\mathrm{p.e.}}$ leads to expectation values of the type
\begin{align}
\Gamma_{k_1,k_2}(t_2,t_1;t_{1'},t_{2'})
= \Braket{c^\dagger_{k_2}(t_2) c^\dagger_{k_1}(t_1) c_{k_1}(t_{1'}) c_{k_2}(t_{2'})},
   \label{Eq4}
\end{align}
where the times are ordered such that $t_1 > t_2$ (see \fb{S.M.} for a detailed derivation) and the spin indices are suppressed for clarity.
Note that this is not a time-ordered expectation value, so the order of the operators remains fixed --- the
unprimed operators come after the primed.  
The total cross section includes the sum of $\Gamma_{k_1,k_2}$ and the same term
with the quasiparticle momenta exchanged, $\Gamma_{k_2,k_1}$.

A useful and intuitive way to visualize two-particle correlation functions $G_2$ (similar to \cref{Eq4}) is diagrammatically, as shown in \cref{fig:pendiagram1}(a). A general vertex describing the propagation of two particles from $(1',2')$ to $(1,2)$ can be drawn with
creation operators as lines
coming out of the vertex, and annihilation operators as
lines going in. They also dictate 4 external legs that then can undergo various complex processes in the shaded square. The power of such $G_2$ diagrams is that they directly describe scattering processes similar to the ones discussed within our paper. Within the shaded square region of the generic diagram, complex interactions can occur, highlighting the power of such $G_2$ diagrams to be adapted for directly describing various scattering processes, as detailed below. 

Without introducing any further interactions (\textit{e.g.} free electrons),
and as long as momentum is a good quantum number, the only 
way to connect this diagram with bare fermion
propagators is shown in \cref{fig:pendiagram1}(b).
This yields a
simple factorized form 
\begin{align}
   \Gamma_{k_1,k_2}(t_2,t_1;t_{1'},t_{2'})
   &=
    \langle c_{k_2}^\dagger(t_2) c_{k_2}(t_{2'}) \rangle  \langle c_{k_1}^\dagger(t_1) c_{k_1}(t_{1'}) \rangle   \nonumber \\
   & = -G_{k_2}^<(t_{2'},t_2) G_{k_1}^<(t_{1'},t_1).
\end{align}
This is just the product of two independent ARPES signals,
which is a natural piece of the \CARPES process but yields no additional
information beyond standard ARPES: the lesser Green's function in frequency domain is simply related to the
spectrum by the fluctuation-dissipation relation
\begin{align}
    G_\kk^<(\omega) = -2i n_F(\omega) \mathrm{Im} G_\kk^R(\omega),
\end{align}
where $n_F(\omega)$ is the Fermi function.
This result is also clear from the diagrammatic structure,
which consists of two fully independent electron propagators [c.f. \cref{fig:pendiagram1}(b)]. This diagram and the corresponding term 
captures the background that arises from the two independent single-electron ARPES
processes, and it is identified as the uncorrelated $\UU$ signal. The additional diagrams we will consider are part of the
correlation signal.

\subsection{Interacting electrons}

A more interesting case arises when an additional interaction is introduced;
this is most easily handled in diagrammatic form. 
The first term that does
not result in a renormalization of the independent single propagators
is shown in \cref{fig:pendiagram1}(c), where a momentum $\qq$ is
transferred from one of the electrons to the other. For example,
if the interaction is an unscreened Coulomb interaction $H_C$, this would
correspond to a correlation function
\begin{align}
\Gamma_{k_1,k_2} & (t_2,t_1; t_{1'},t_{2'})
= \nonumber \\
&\int dt_3
\Braket{c^\dagger_{k_2}(t_2) c^\dagger_{k_1}(t_1) H_C(t_3)c_{k_1}(t_{1'}) c_{k_2}(t_{2'})}
    \label{eq:firstordercoulomb}
    \\
    H_{\mathrm{C}} &=  \sum_{\qq\ell\ell'} V_\qq 
    c^\dagger_{\ell'-\qq} c^\dagger_{\ell+\qq} c_{\ell} c_{\ell'}.
\end{align}
where the internal $t_3$ variable is understood to occur between external
times. $H_C$ is placed in the center the expression found in the previous section and with the assignment of $(1,1',2,2')$ as the external legs due to the time-ordering of the integration limits in the full form.
This example illustrates the use of  \CARPES for measuring interactions between
electrons. The Wick contraction
of \cref{eq:firstordercoulomb} matches the momenta of the external 
electrons ($\kone$ and $\ktwo$, which are experimentally determined)
with the momenta involved in the interactions. In other words,
the experiment probes the amplitude of the interaction $V_{\qq}$ at $\qq = k_2 - k_1$.

\begin{figure}[h]
    \includegraphics[width=0.4\textwidth]{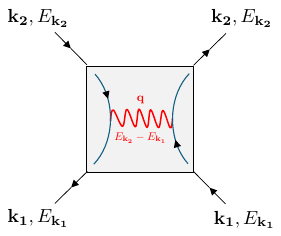}
    \caption{Feynman diagram $G_2$ in energy and momentum space for the the  \CARPES scattering process with a single scattering event that transfers momentum $\qq$
    and energy $E_\kone - E_{\ktwo}$. The momenta and energies indicated
    are those of the photoemitted electrons 
    measured by the  \CARPES experiment.}
    \label{fig:pen_Vqw}
\end{figure}

This idea extends to dynamic interactions, \emph{i.e.} that depend on both energy and momentum.
Working in the limit of a long pulse --- i.e. where the
duration of the probe pulse is much longer than the
characteristic time scale associated with the energies
involved ---
we can effectively restore time translation invariance
and can convert the time domain diagram
of \cref{fig:pendiagram1}(c) into the frequency domain version shown in \cref{fig:pen_Vqw}.
A straightforward evaluation this diagram, which will appear as a
correlation on top of the independent single electron ARPES processes,
yields
\begin{align}
    \CC=\bigg| G^<_\kone(E_\kone) G^<_{\ktwo}(E_{\ktwo})
     \bigg|^2
     \mathrm{Re}\left[V_\qq(E_{\ktwo} - E_\kone)\right] \delta_{q,k_2-k_1}
     \label{eq:ncarpes_vqw}
\end{align}
for the \CARPES vertex, thus
enabling direct access to the interaction
$V_\qq(\omega)$. Only the real part of $V_\qq(\omega)$ appears since the diagram
is symmetric under the exchange of electrons, and
the imaginary part of the interaction is anti-symmetric
in energy.
Note that in \cref{eq:ncarpes_vqw}
the two electron momenta $k_1$ and $k_2$, as well
as the electron energies $E_{k_1}$ and $E_{k_2}$,
are free, but the signal will rapidly decay once the
electrons are not on-shell, \emph{e.g.} when $E_\kone $ is not
close to the electron dispersion $\varepsilon(\kone)$.

Interestingly, the structure of this diagram is quite
close to that for RIXS where the interaction is mediated
by a boson\cite{devereaux2016directly}.  However, the added advantage
of  \CARPES over RIXS is that, through the electron energy and momenta,
one can directly interrogate the interaction of the (possibly effective) boson with \eh{specific}
electron states. This is not possible in the photon-in/photon-out setup
of RIXS, where the signal is averaged over the entire Brillouin zone.

\subsubsection*{Peierls instability}
\label{sec:peierls}

In the following, we illustrate how \CARPES can probe electron-electron interactions using the example of a CDW instability in 1D induced by the Kohn anomaly.
Specifically, we use a parabolic dispersion
$\varepsilon(k)=k^2$ and a Fermi energy $\varepsilon_F=1$.
The nesting of the Fermi surface
causes strong charge fluctuations at momenta $q=2k_F$, where
$k_F$ is the Fermi wavevector; these fluctuations soften the 
phonons at the same momenta.
In this case, the boson
line in \cref{fig:pen_Vqw} is the renormalized phonon,
and we can use  \CARPES to
retrieve the phonon-mediated electron-electron interactions 
$V_\qq(\omega)=g_q^2 D_\qq(\omega)$. We use a deformation potential interaction
vertex $g_q \sim |q|$ and an acoustic phonon with sound velocity $c$ (S.M. section C provides 
details of the Kohn anomaly mechanism).

\begin{figure}
    \centering
    \includegraphics[width=0.99\linewidth]{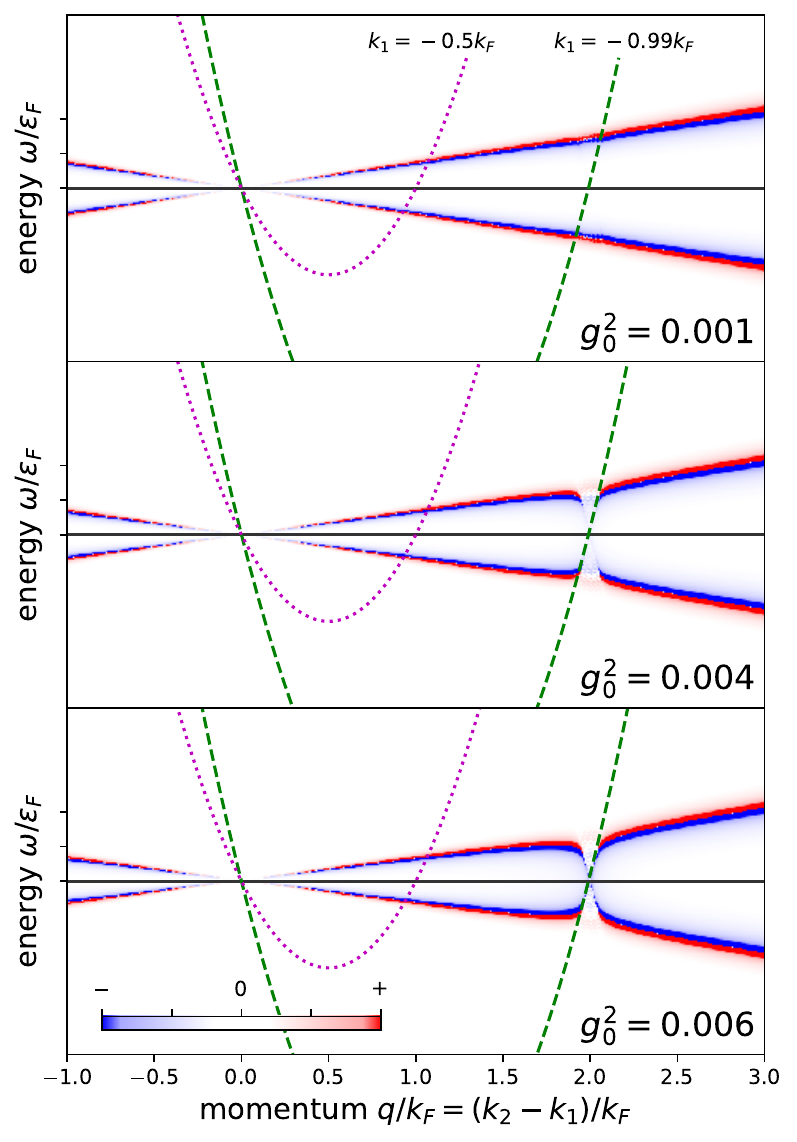}
    \caption{Phonon-mediated electron-electron interaction 
    Re[$V_\qq(\omega)$] shown in false color.
    The dotted (dashed) line are the curves traced out in the $q=k_2-k_1,\omega=E_2-E_1$ plane obtained by fixing $k_1$
    at the indicated value
    and varying $k_2$ with both energies $\varepsilon_{1,2}$ remaining
    on-shell [$E_{1,2} = \varepsilon(k_{1,2})$].}
    \label{fig:D}
\end{figure}
As the electron-phonon coupling strength $g_0$ increases, the peak in the charge
susceptibility (Lindhard function) grows\cite{mihaila2011lindhard}, and the phonon
softens at $q=2k_F$.
Now,  \CARPES probes the real part of the interactions,
displayed in \cref{fig:D} in false color,
by picking out the two photoelectron momenta
and energies. To illustrate how different sets of pairs $(k_1,E_1)$ and $(k_2,E_2)$ select the momentum and energy
along which we measure $V_\qq$ via C-ARPES we use non-interacting bands,
which fix the energies to be
equal to the on-shell energies [\emph{i.e.} simply on the electron
band $\varepsilon(k)$], so that the only two free variables are the photoelectron
momenta $k_1$ and $k_2$. We next pick $k_1=-0.99k_F$ 
such that $\varepsilon(k_1)$ lies just below the Fermi level for left-moving electrons, and consider the  \CARPES signal as a function of $k_2$. This traces out the dashed green curve
shown in \cref{fig:D}, and measures the electron-phonon interaction along
this line. Choosing a different momentum such as $k_1=-0.5k_F$ traces out a
different slice of the interactions (dotted purple line).

\begin{figure}[htpb]
    \centering
    \includegraphics[width=0.99\linewidth]{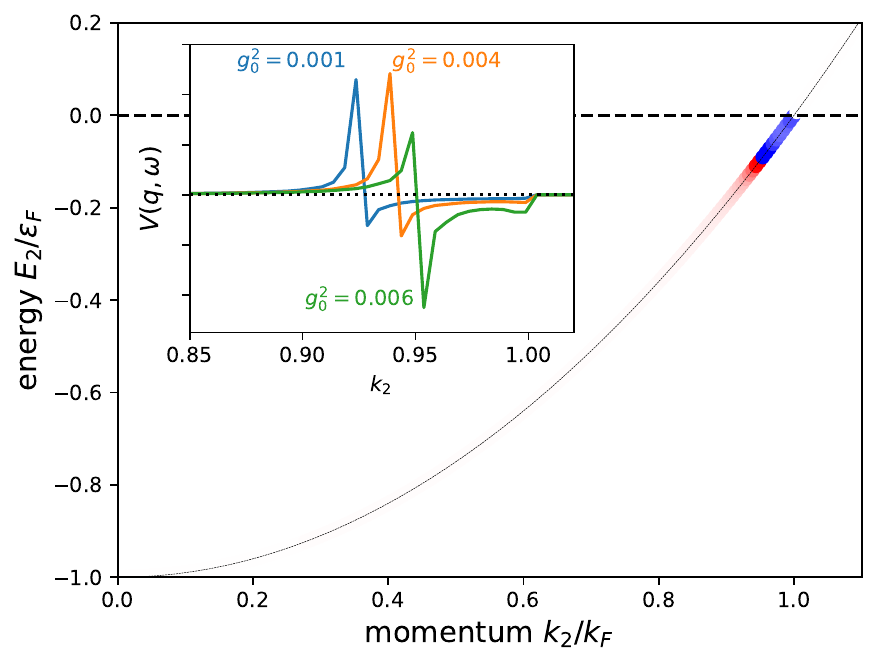}
    \caption{Phonon mediated electron-electron interaction $V_\qq(\omega)=g_0^2 D_\qq(\omega)$ as measured
    with  \CARPES. The main panel shows the result of an experiment where
    $k_1=-0.99k_F$, and $k_2$ is varied. A nonzero signal is only observed along the band dispersion (see text). Inset: Signal as measured along the
    band for various values of the coupling strength $g_0$.}
    \label{fig:Vqw_onshell}
\end{figure}

\begin{figure*}[ht]
    \includegraphics[width=1\linewidth]{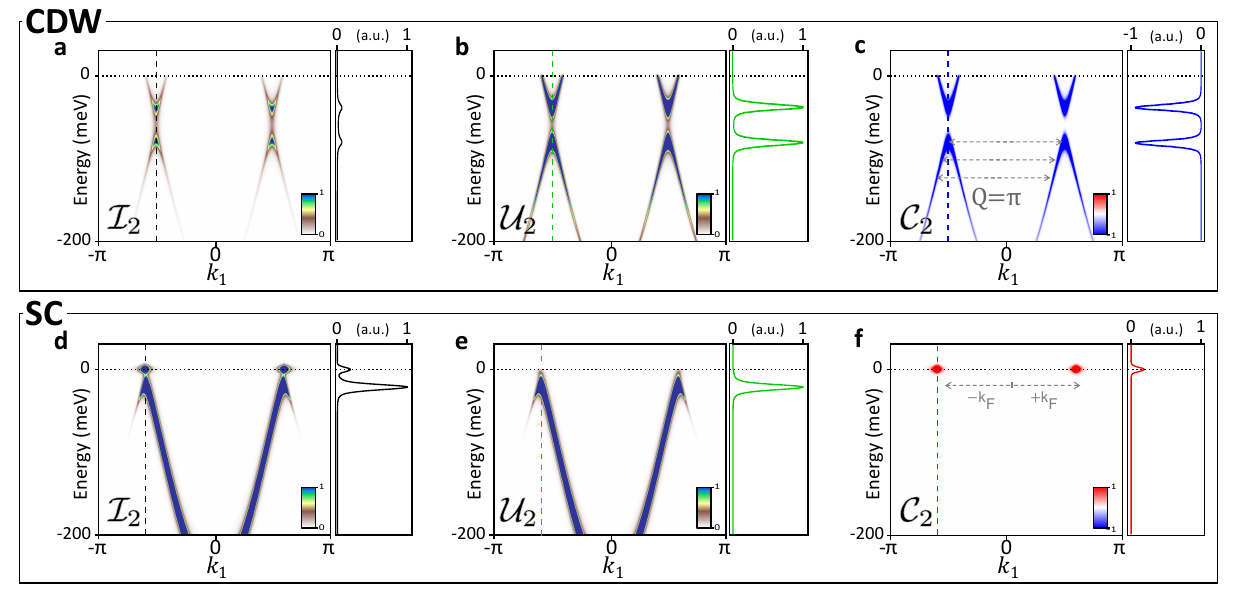}
    \caption{ Total coincidence signal (\(\II\)), uncorrelated background (\(\UU\)), and \CARPES intensity (\(\CC\)), considering all possible electrons with momentum $k_1$ for a 1D CDW system, with $\ktwo=\kone+Q$ (a-c), and for a SC, with $\ktwo=-\kone$ (d-f). Here we use $\Delta=20$ meV for both CDW and SC gaps, and a broadening in energy of 5 meV is used. Energy distribution curves (on the right) are plotted to highlight the contribution of correlations to \(\II\).  }
    \label{fig:1D-SC-CDW}
\end{figure*}

\Cref{fig:Vqw_onshell} displays the result of such an experiment,
where we have fixed $k_1=-0.99k_F$. The majority of the signal occurs
when the second photoelectron is also emitted from near the Fermi level.
Moreover, as the interaction strength $g_0$ increases the interaction $V_\qq(\omega)$
becomes more attractive (negative Re[$V_\qq(\omega)$]) which eventually leads to the formation of a CDW\cite{gruner1988dynamics}.

It is interesting to note that at $k_1=-0.5k_F$, there is also a $k_2$ where
a nonzero signal can be observed, even though this is not at the typical
$q=2k_F$ that is commonly associated with CDW formation in 1D.
This can be seen in \cref{fig:D} where the dotted line intersects a non-zero part
of $V_\qq(\omega)$ away from $\omega=0$.
The signal
appears at  $k_2 < 2k_F$, where the electron dispersion intersects with the
sharp feature in the phonon-mediated interaction at $\Omega=E_2-E_1 \approx cq$, where $c$ is the sound velocity. 
In other words, C-ARPES can be employed to map the dispersion of the bosonic mode.
This example clearly underscores the capability of C-ARPES in elucidating the interactions that ultimately give rise to the emergent CDW phase, demonstrating its potential to precisely identify the electrons involved in the process.

\subsection{Ordered Phases}
Having discussed the extraction of dynamic electron interactions via \CARPES, we present two key examples of the \CARPES signal for ordered phases in 1D: CDW and superconductivity (SC).
There are clear distinctions between the two cases, which can already be seen by considering the instantaneous correlations. For SC, Stahl et al.\cite{stahl2019noise} show that for an ideal ultrashort pulse $\CC \propto |\langle c_{k\uparrow}c_{-k\downarrow}\rangle|^2$.  For a CDW, a similar calculation yields $\CC \propto -|\langle c^\dagger_k c^{\phantom{\dagger}}_{k+Q}\rangle |^2$, which has an opposite sign. This is a qualitative distinction that arises in addition to the momentum and energy structure that we discuss in this section.

\subsubsection{Charge density wave}

We consider the case of a simple CDW with wavevector $Q$ that is described by an \emph{anomalous} off-diagonal
Green's function, such as $\langle c^\dagger_{k+Q}(t) c_k(t')\rangle$. With this possibility, the two points of
non-equal momenta in the \LK{$G_2$} diagram can be directly connected
with $k_2=k+Q$.
We can read off from the diagram of \cref{fig:pendiagram1}(d) that the  \CARPES correlation function, on top of the uncorrelated contribution $\UU$, is 
\begin{align}
\Gamma_{\kone,\ktwo}& =
    \langle c^\dagger_{\kone+Q}(t_2) c_\kone(t_{1'}) \rangle 
    \langle c^\dagger_\kone(t_1) c_{\kone+Q}(t_{2'}) \rangle 
     \\
    & = -F_\kone^<(t_{1'},t_2) F_{\kone+Q}^<(t_{2'},t_1),
\end{align}
where $F_\kone$ and $F_{\kone+Q}$ denote the two off-diagonal
components of the $2\times 2$ Nambu Green's function $\mathcal{G}_{k_1}$.
In the frequency domain, this becomes
\begin{align}
\CC = 
    -F_\kone^<(E_\kone) F_{\kone+Q}^<(E_{\kone+Q}) = -\Big|F_\kone^<(E_\kone)\Big|^2
\end{align}
having used the additional constraint that the photoelectron energies
are equal, $E_\kone = E_{\kone+Q}$ (\emph{i.e.}, static CDW order). 
We note that
the same result can be obtained
from straightforward perturbation theory, without referring to diagrams.

The retarded Green's function for a mean field CDW is 
\begin{align}
\mathcal{G}_k^R(\omega) = \frac{1}{(\omega - \xi_+ + i0^+)\tau_0 - \xi_- \tau_3 - \Delta \tau_1}.
\label{eq:G}
\end{align}
Here, $\tau_i$ are the Pauli matrices, 
$\xi_{\pm} = \varepsilon(\kk) \pm \varepsilon(\kk+Q)$,
where $\varepsilon(\kk)$ is the bare quasiparticle dispersion, and $\Delta$ is the CDW gap.
The anomalous propagator $F$ is the off-diagonal component of
this matrix.
Via the fluctuation-dissipation relation, the \CARPES signal
is then
\begin{align}
\CC = -
    \Bigg| 2 n_F(E_\kone) \mathrm{Im} \frac{\Delta}{(E_\kone - \xi_+ + i0^+)^2 - (\xi_-^2 + \Delta^2)} \Bigg|^2,
    \label{eq:FF_CDW}
\end{align}
where $n_F$ is the Fermi-Dirac distribution.

We now select the two photoelectrons to have the same energy $(E_\kone = E_\ktwo)$, and
fix their momenta to be separated by the CDW wavevector $Q$.
In the non-interacting limit, the four free variables ($\kone, \ktwo, E_\kone, E_\ktwo$) are reduced to just two
by the relationships $\ktwo = \kone + Q$ and $E_\kone = E_\ktwo$, and in what follows we consider \CARPES
as a function of these two.
In \cref{fig:1D-SC-CDW}, in the top row, we show the results obtained for a 1D CDW system as a function of the
remaining free variable $\kone$ and $E_\kone$ for a mean field CDW with a finite chemical potential, which shifts the
gap away from the Fermi level. 
\cref{fig:1D-SC-CDW}c shows the \CARPES intensity \(\CC\), which is the difference between the coincidence signal $\II$ (panel a) and the
uncorrelated single electron background \(\UU\) (panel b). 
We observe a strong negative \CARPES intensity in the vicinity of the single particle spectral gap, some distance below the Fermi level.
The energy distribution curve at $\kone$, where the CDW gap opens, is plotted on the side of each panel and clearly
highlights the strong suppression observed for the total coincidence intensity $\II$, given by the negative $\CC$, if compared with the uncorrelated contribution $\UU$.

\subsubsection{Superconductivity} 

The superconducting case has been already discussed in detail by
Stahl et al.\cite{stahl2019noise}. The key difference from the CDW case
is that the time-ordered versions arise instead of occupied (lesser)
Green's functions, which contain the square modulus of the imaginary part of the spectrum.
This fact, together with the symmetric nature of the
superconducting spectrum, limits the signal to only
being visible near $E_k=E_{-k}\approx 0$ (see also
Refs.~\cite{mahmood2022distinguishing,devereaux2023angle,su2020coincidence}).
The resulting
\CARPES signal for a 1D SC system is
\begin{align}
    \CC=\Bigg| \frac{\Delta}{(\omega + i0^+)^2 - (\xi_k^2 + \Delta^2)} \Bigg|^2 \cdot \delta(E_\kk + E_{-\kk}).
    \label{eq:FF_SC}
\end{align}

In \cref{fig:1D-SC-CDW}, in the bottom row, $\CC$, $\II$ and $\UU$ are evaluated in the case of a 1D BCS-like superconductor ($\Delta$= 20meV), for any possible $k_1$ electron interacting with a $k_2=-k_1$ electron at the same energy. 
On the right side of panels d-f, the energy distribution curve at the Fermi wavevector is shown for direct evaluation of the  \CARPES intensity. The  \CARPES signal, $\CC$, is characterized by a peak at the Fermi energy that has no counterpart in the ARPES spectrum, demonstrating the potential of \CARPES to explore the properties of correlated systems with unprecedented direct access to the two-electron correlation function with momentum resolution.

\section{Discussion}

\CARPES promises an unprecedented ability to untangle the otherwise averaged interactions by picking out specific electron pairs and examining their interactions. In this work, 
we have applied this to the investigation of a phonon-mediated
effective electron-electron interaction,
and we have just begun to lay the foundation for the \CARPES technique by considering mean-field CDW and SC orders in 1D, with the pairing function
peaked at a singular wavevector and without energy dependence. 
However, \CARPES could help us to go beyond our current understanding of quantum materials by directly studying the energy dependence of the effective pairing boson underlying different emergent phases, such as strong-coupling Eliashberg superconductors\cite{eliashberg1960interactions,scalapino1966strong,agd} and fluctuating charge density wave orders\cite{kivelson2003detect,wandel2022enhanced,daSilvaNeto_Frano_Boschini_Frontiers_RIXS_review}.

Within the context of charge density waves, there is uncertainty regarding the mechanism by which they are formed
that could be addressed with C-ARPES.
Specifically, both electron-lattice (Peierls) and electron-electron (Coulomb) interactions could give rise to charge density wave instabilities\cite{rossnagel2011origin,watanabe2015charge,van2018competing}.
Both mechanisms involve the complex interplay of Fermi surface topology, momentum-dependence of the electron-phonon vertex\cite{eiter2013alternative}, and other details that go into determining the effective interaction.
For example, the random phase approximation (RPA), commonly employed to simulate and compare with scattering experiments, relies on both the polarizability function and the effective potential. The standard Lindhard equation for the polarizability involves summing over all possible electronic states. However, this approach often fails to quantitatively capture certain scattering experiments, 
possibly because it neglects an unknown vertex that weights
the momenta that go into the calculation. 
Indeed, in strongly correlated systems with charge order, such as cuprate high-$T_c$ superconductors, the RPA breaks down, as evidenced by momentum-resolved electron energy loss spectroscopy (M-EELS) and RIXS experiments \cite{Mitrano_PNAS, Hussain_PRX_2019, CO_rings}. 

In a similar manner to the approach discussed in Section II.B in the context of Peierls instabilities,
C-ARPES could also be employed to resolve the $\Omega-q$ structure of bosonic modes in situations where they are inaccessible through direct scattering experiments.
For example, also in the high-$T_c$ cuprates
there remains uncertainty about the nature of the dynamic correlations associated with this charge order. Based on anomalies observed in the RIXS-measured phonon line, an acoustic low-energy dispersion was proposed \cite{Chaix_CO_BSCCO, Li_PNAS_2020, CO_melting}. However, extending measurements beyond the high-symmetry directions of the Brillouin zone reveals a more intricate structure of dynamic charge order correlations, likely reflecting the effective form of the Coulomb interactions \cite{CO_rings, QCDCs_phonon, daSilvaNeto_Frano_Boschini_Frontiers_RIXS_review}.

Finally, 
while the present study focused on the theoretical modeling of the \CARPES signal, it is nonetheless essential to briefly address the relationship between the theoretical framework outlined herein and its practical experimental implementation. \CARPES, akin to 2e-ARPES, relies on the coincident detection of two electrons, highlighting the need for a specialized coincidence electron detection setup. This may entail the implementation of two synchronized conventional single-hit electron detectors \cite{zwettler2024extreme,chiang2020laser,trutzschler2017band,huth2014electron} or a specialized multi-hit electron detector \cite{wallauer2012momentum,voss2020time}. 

In the specific case of \CARPES, it is evident that the temporal and spatial envelope of the UV beam should be of the order of the underlying $\Gamma_{k_1,k_2}$ correlation length and time (for example, in conventional superconductors, these two are typically on the order of tens-to-hundreds of femtoseconds and nanometers, respectively). In scenarios where the temporal and spatial profiles of a UV pulse exceed the intrinsic coherence time and length of electron correlations and/or in the presence of non-ideal detection schemes and extrinsic background sources, the $\CC$-signal-to-background ratio may decrease.

In experimental \CARPES efforts, by continuously changing the temporal duration or the spot size of the UV beam, we will effectively separate the emergence of the $\CC$ signal from the underlying background. We believe that because of its favorable cross-section and direct access to $\Gamma_{k_1,k_2}$, rather than a higher order scattering process such as is involved in 2e-ARPES
where a single photon is responsible for emitting two electrons \cite{mahmood2022distinguishing,devereaux2023angle,chiang2020laser,zwettler2024extreme},  \CARPES has the potential to unravel the interactions into resolvable, understandable building blocks upon which theories can be constructed and tested.

\section*{Author contributions}
AFK, EHdSN and FB conceived the original idea for the work.
AFK and HL developed the theoretical framework of the CARPES response. 
FG contributed to the analysis of mean field CDW and SC order.
FG, NG, EHdSN, and FB provided the experimental context.
All authors contributed to the writing of the manuscript, and read and approved the final manuscript.

\begin{acknowledgments}
We acknowledge helpful discussions with Steve Johnston.
This research is funded by the Gordon and Betty Moore Foundation’s EPiQS Initiative, Grant GBMF12761 to F.B., A.F.K., and E.H.d.S.N. F.B. acknowledges support from the Natural Sciences and Engineering Research Council of Canada (NSERC), the Canada Research Chairs Program, 
the Canada Foundation for Innovation (CFI); the Fonds de recherche du Qu\'{e}bec --- Nature et Technologies (FRQNT), and the Minist\`{e}re de l'\'{E}conomie, de l'Innovation et de l'\'{E}nergie --- Qu\'{e}bec.
\end{acknowledgments}

\clearpage
\onecolumngrid
\appendix

\renewcommand\thefigure{S\arabic{figure}}  
\renewcommand\thetable{S\arabic{table}}  
\setcounter{figure}{0}

\renewcommand{\kone}{{k_1}}
\renewcommand{\ktwo}{{k_2}}

\section{Technical details}
\label{app:derivation}
In this section, we derive the diagrammatic representation of the  \CARPES process. We begin from the 
expression obtained by Stahl et al.\cite{stahl2019noise} (who termed it noise-correlation ARPES)
for the 2-particle signal $\II$.  Note that here we explicitly distinguish between
the ejected photoelectron momenta (denoted with $p$) and the quasiparticle momenta (denoted with $k$).
\begin{align}
\II =
\sum_{\substack{k_1,k_2,k_{1'},k_{2'}\\\sigma_1,\sigma_2,\sigma_{1'},\sigma_{2'}}}
&\int_{-\infty}^\infty dt_1
\int_{-\infty}^{t_1}  dt_2
\int_{-\infty}^{\infty} dt_{1'}
\int_{-\infty}^{t_{1'}}  dt_{2'}
S(t_1)^\ast S(t_2)^\ast S(t_{1'})S(t_{2'})
e^{-\imag\Omega(t_{1'}+t_{2'}-t_1-t_2)}\nonumber\\
\times&\Big[
(M_{k_1,p'}^{\sigma_1,\tau'})^\ast
(M_{k_2,p}^{\sigma_2,\tau})^\ast
M_{k_{2'},p}^{\sigma_{2'},\tau}
M_{k_{1'},p'}^{\sigma_{1'},\tau'}\Braket{c^\dagger(2) c^\dagger(1)c(1')c(2')}_0^c e^{-\imag E_p(t_2-{t_{2'}})}e^{-\imag E_{p'}(t_1-t_{1'})}\nonumber
 \\
&+
(M_{k_1,p}^{\sigma_1,\tau})^\ast
(M_{k_2,p'}^{\sigma_2,\tau'})^\ast
M_{k_{2'},p'}^{\sigma_{2'},\tau'}
M_{k_{1'},p}^{\sigma_{1'},\tau}\Braket{c^\dagger(2) c^\dagger(1)c(1')c(2')}_0^ce^{-\imag E_p(t_1-{t_{1'}})}e^{-\imag E_{p'}(t_2-t_{2'})}
\label{eq:stahlApp}
\\
&-(M_{k_1,p'}^{\sigma_1,\tau'})^\ast
(M_{k_2,p}^{\sigma_2,\tau})^\ast
M_{k_{2'},p'}^{\sigma_{2'},\tau'}
M_{k_{1'},p}^{\sigma_{1'},\tau}
\Braket{c^\dagger(2) c^\dagger(1)c(1')c(2')}_0^c e^{-\imag E_p(t_2-{t_{1'}})}e^{-\imag E_{p'}(t_1-t_{2'})}\nonumber
\\
&-(M_{k_1,p}^{\sigma_1,\tau})^\ast
(M_{k_2,p'}^{\sigma_2,\tau'})^\ast
M_{k_{2'},p}^{\sigma_{2'},\tau}
M_{k_{1'},p'}^{\sigma_{1'},\tau'}
\Braket{c^\dagger(2) c^\dagger(1)c(1')c(2')}_0^c e^{-\imag E_p(t_1-{t_{2'}})}e^{-\imag E_{p'}(t_2-t_{1'})}\Big].\nonumber
\end{align}
Note that here, $1 = (t_1, k_1, \sigma_1)$. The $S(t)$ functions are shape functions that are set by the probe profile;
the $M^{\sigma_1,\tau}_{k,p}$ factors indicate the photoemission matrix element for a quasiparticle 
with momentum $k$ and spin $\sigma$ becoming a free electron with momentum $p$ and spin $\tau$;
$E_p$ denotes the energy of the free photoemitted electron with momentum $p$.  In what follows, we will suppress
the spin indices --- these can be straightforwardly restored. We will also assume the matrix elements conserve spin
and momentum, i.e. $M^{\sigma,\tau}_{k,p} = M \delta_{\sigma,\tau} \delta_{k,p}$.

\begin{figure}
    \includegraphics[width=0.6\textwidth]{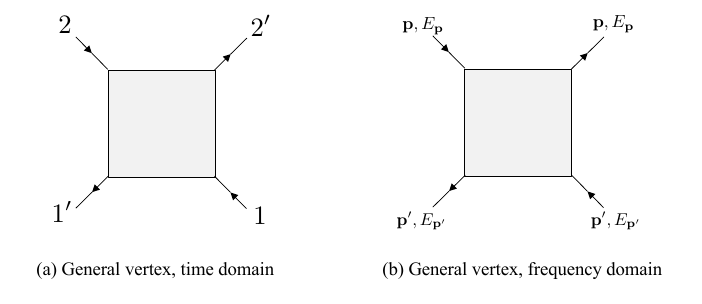}
    \caption{Feynman diagram illustrating the 4-point vertex for the scattering process, where the primed (nonprimed) times and 4-momenta correspond to creation (annihilation) operators and
    have been drawn as outgoing (incoming) lines.}
    \label{fig:pendiagram_appendix}
\end{figure}

\Cref{eq:stahlApp} can be significantly simplified by making two observations.  First: the first and second lines are identical
under the exchange of the free electron momenta $p \leftrightarrow p'$ (even without the matrix element simplification), and the same holds for the third and fourth lines.
This yields
\begin{align}
\II = |M|^4
&\int_{-\infty}^\infty dt_1
\int_{-\infty}^{t_1}  dt_2
\int_{-\infty}^{\infty} dt_{1'}
\int_{-\infty}^{t_{1'}}  dt_{2'}
S(t_1)^\ast S(t_2)^\ast S(t_{1'})S(t_{2'})
e^{-\imag\Omega(t_{1'}+t_{2'}-t_1-t_2)}\nonumber\\
\times&\Big[
\Braket{c_p^\dagger(t_2) c_{p'}^\dagger(t_1)c_{p'}(t_{1'})c_p(t_{2'})}_0^c e^{-\imag E_p(t_2-{t_{2'}})}e^{-\imag E_{p'}(t_1-t_{1'})}\nonumber
 \\
&-
\Braket{c_p^\dagger(t_2) c_{p'}^\dagger(t_1)c_{p'}(t_{1'})c_p(t_{2'})}_0^c e^{-\imag E_p(t_2-{t_{1'}})}e^{-\imag E_{p'}(t_1-t_{2'})}
\Big]
\label{eq:stahl_app2}
\\
&+ p \leftrightarrow p'  \nonumber
\end{align}
where we have explicitly denoted the time and momentum arguments. Our second observation is that we can exchange
the dummy arguments
$t_{1'}$ and $t_{2'}$ in the second line, and restore their ordering by swapping the operators,
\begin{align}
\II = |M|^4
&\int_{-\infty}^\infty dt_1
\int_{-\infty}^{t_1}  dt_2
\int_{-\infty}^{\infty} dt_{1'}
\int_{-\infty}^{\infty}  dt_{2'}
S(t_1)^\ast S(t_2)^\ast S(t_{1'})S(t_{2'})
e^{-\imag\Omega(t_{1'}+t_{2'}-t_1-t_2)}\nonumber\\
\times&
\Braket{c_p^\dagger(t_2) c_{p'}^\dagger(t_1)c_{p'}(t_{1'})c_p(t_{2'})}_0^c e^{-\imag E_p(t_2-{t_{2'}})}e^{-\imag E_{p'}(t_1-t_{1'})}
\\
&+ p \leftrightarrow p'  \nonumber
\label{eq:stahl_app3}
\end{align}
Note that the upper limit of the $t_{2'}$ integration is now extended to $\infty$. In what follows, we will make use of a general vertex $\Gamma_{p,p'}(t_2,t_1;t_{1'},t_{2'})$, or in the frequency domain, $\tilde\Gamma_{p,p'}(\alpha_2, \alpha_1; \alpha_{1'},\alpha_{2'})$, represented by \cref{fig:pendiagram1} which we reproduce in \cref{fig:pendiagram_appendix} with slight modification to match the notation.
With this,  we can express $\II$ as
\begin{align}
\II = |M|^4
	&\int_{-\infty}^\infty dt_1
	\int_{-\infty}^{t_1}  dt_2
	\int_{-\infty}^{\infty} dt_{1'}
	\int_{-\infty}^{\infty}  dt_{2'}
	S(t_1)^\ast S(t_2)^\ast S(t_{1'})S(t_{2'})
	e^{-\imag\Omega(t_{1'}+t_{2'}-t_1-t_2)}\\
	\times&
	\Gamma_{p,p'}(t_2,t_1;t_{1'},t_{2'}) \cdot
	e^{-\imag E_p(t_2-{t_{2'}})}e^{-\imag E_{p'}(t_1-t_{1'})} \nonumber
\\
&+ p \leftrightarrow p'  \nonumber \\
= |M|^4
	&\int_{-\infty}^\infty dt_1
	\int_{-\infty}^{t_1}  dt_2
	\int_{-\infty}^{\infty} dt_{1'}
	\int_{-\infty}^{\infty}  dt_{2'}
	S(t_1)^\ast S(t_2)^\ast S(t_{1'})S(t_{2'})
	e^{-\imag\Omega(t_{1'}+t_{2'}-t_1-t_2)}\\
	\times&
	\left[ \int d\alpha_1 d\alpha_2 d\alpha_{1'} d\alpha_{2'}
	\Gamma_{p,p'}(\alpha_2, \alpha_1; \alpha_{1'},\alpha_{2'})
	e^{i \alpha_2 t_2}
	e^{i \alpha_1 t_1}
	e^{-i \alpha_{2'} t_{2'}}
	e^{-i \alpha_{1'} t_{1'}}
	\right]
	\cdot
	e^{-\imag E_p(t_2-{t_{2'}})}e^{-\imag E_{p'}(t_1-t_{1'})} \nonumber
\\
&+ p \leftrightarrow p'   \nonumber
\end{align}
where we have implicitly used time-translation invariance to express the vertex in the frequency domain.  
We can work out the frequency (time) integrals over the $\alpha$ ($t$) variables; the shape functions here are used to 
regularize the Fourier integrals, and we will take the limit of $S(t) \rightarrow 1$ at the end.  We find
\begin{align}
\II = \frac{|M|^4}{2} 
&
	\int d\alpha_1 d\alpha_2 d\alpha_{1'} d\alpha_{2'}\,
	\Gamma_{p,p'}(\alpha_2, \alpha_1; \alpha_{1'},\alpha_{2'}) \nonumber\\
	&\times
	\delta(E_p - \Omega - \alpha_{1'}) \cdot
	\delta(E_{p'} - \Omega - \alpha_{2'}) 
	\label{eq:frequency_vertex} \\
	&\times
	\delta\left[ \alpha_1 + \alpha_2 - \alpha_{1'} - \alpha_{2'} \right] \cdot
	\delta\left[ (\alpha_1 - \alpha_2) - (E_p - E_{p'}) \right].\nonumber\\
	& + p \leftrightarrow p'  \nonumber \\
	= \frac{|M|^4}{2} 
    &
	\Gamma_{p,p'}(E_{p'}-\Omega, E_p-\Omega; E_p-\Omega , E_{p'} - \Omega) \nonumber\\
	& + p \leftrightarrow p' 
\end{align}
In \cref{eq:frequency_vertex}, the first set of delta functions indicates the relationship between the energy entering the diagram
from the photon and photoemitted electron and the internal variables; the second set speaks to energy conservation
within the diagram. Finally, we can measure the photoelectron energies $E_{p,p'}$ with respect to the
photon energy, shifting them $E_{p,p'} -\Omega \rightarrow E_{p,p'}$.
The resulting frequency domain vertex is shown in \cref{fig:pendiagram_appendix}.

\section{$\II$ for superconductivity}

The case for superconductivity was worked out by Stahl et
al.\cite{stahl2019noise}. Here, it is convenient to re-arrange
the terms in \cref{eq:stahlApp} somewhat differently, leading to
\begin{align}
\II = |M|^4
&\int_{-\infty}^\infty dt_1
\int_{-\infty}^\infty  dt_2
\int_{-\infty}^{\infty} dt_{1'}
\int_{-\infty}^{\infty}  dt_{2'}
S(t_1)^\ast S(t_2)^\ast S(t_{1'})S(t_{2'})
e^{-\imag\Omega(t_{1'}+t_{2'}-t_1-t_2)}\nonumber\\
\times&
\Braket{
T_{\bar{t}}[c_p^\dagger(t_2) c_{p'}^\dagger(t_1)]
T_{t}[c_{p'}(t_{1'})c_p(t_{2'})]
}_0^c e^{-\imag E_p(t_2-{t_{2'}})}e^{-\imag E_{p'}(t_1-t_{1'})}
\\
&+ p \leftrightarrow p'  \nonumber
\end{align}
where $T_t$ ($T_{\bar{t}}$) denotes time-ordering (anti-time ordering) of the terms in brackets.  Performing a Wick
contraction yields for the anomalous term
\begin{align}
\II = |M|^4
&\int_{-\infty}^\infty dt_1
\int_{-\infty}^\infty  dt_2
\int_{-\infty}^{\infty} dt_{1'}
\int_{-\infty}^{\infty}  dt_{2'}
S(t_1)^\ast S(t_2)^\ast S(t_{1'})S(t_{2'})
e^{-\imag\Omega(t_{1'}+t_{2'}-t_1-t_2)}\nonumber\\
\times&
(F^\dagger_p)^{\bar{t}}(t_2-t_1)
(F_p)^t(t_2'-t_1')
e^{-\imag E_p(t_2-{t_{2'}})}e^{-\imag E_{-p}(t_1-t_{1'})}
\\
&+ p \leftrightarrow p'  \nonumber
\end{align}

\section{Phonon mediated electron-electron scattering}
\label{app:peierls}
\begin{figure}[htpb]
    \includegraphics[width=0.4\textwidth]{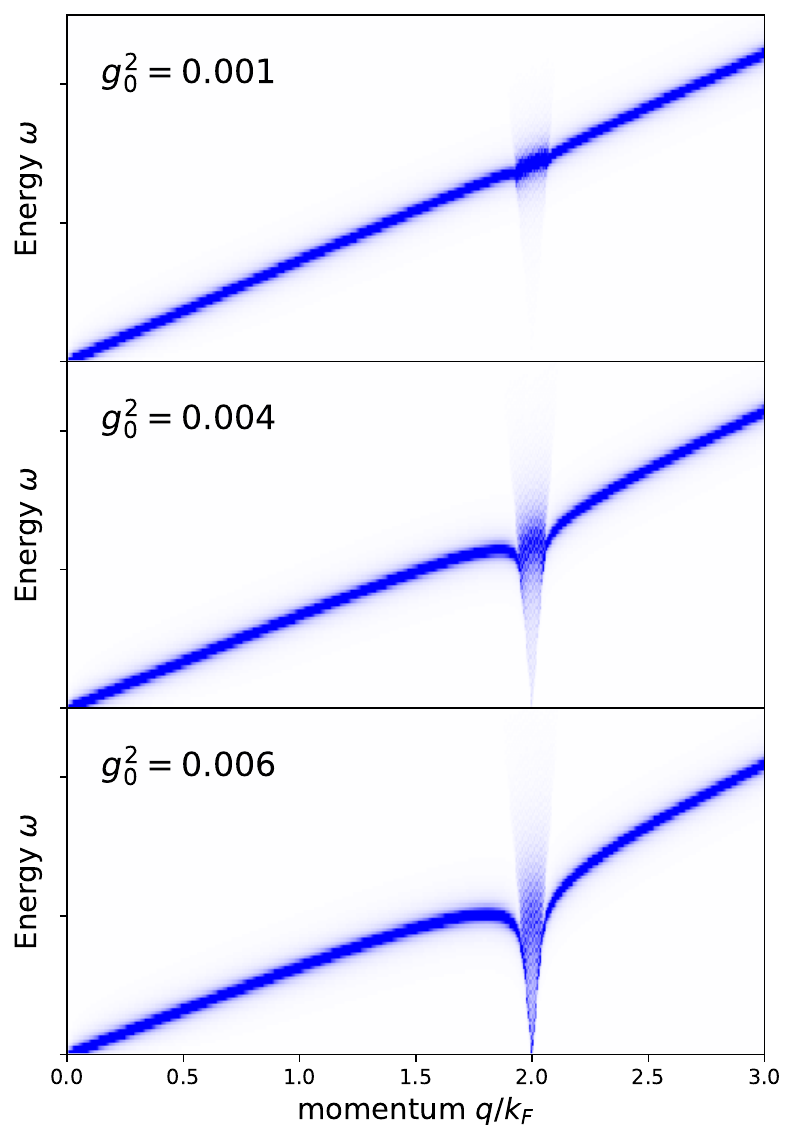}
    \caption{Phonon spectral function $(-1/\pi)\mathrm{Im} D(\qq,\omega)$ showing the development
    of the Kohn anomaly at $q=2k_F$.}
    \label{fig:kohn_D}
\end{figure}

In Sec.~\ref{sec:peierls}, we illustrate the use of  \CARPES for measuring the interaction between electrons with the example of phonon-mediated
scattering; in this section, we provide some additional details,
which can be found in standard many-body theory textbooks\cite{Mahan, bruus}.
We consider the interaction between electrons and phonons for a
single longitudinal acoustic phonon in 1D
\begin{align}
    H_{\mathrm{ep}} &= \frac{1}{\sqrt{V}}
        \sum_{\qq\kk} g_\qq c^\dagger_{\kk+\qq} c_\kk \left[ a_\qq + a^\dagger_{-\qq} \right], \\
        g_\qq &= g_0 \frac{|\qq|^2}{\Omega_q}, \\
        \Omega_\qq &= c |\qq|.
\end{align}
In metals, phonons exhibit the Kohn anomaly, where there is a
divergent response of the electron polarization $\Pi(\qq,\omega)$
at a momentum $q=2k_F$. This quantity enters the phonon
propagator via the Dyson equation
\begin{align}
D(\qq,\omega) = 
    \frac{2\Omega_\qq}{\omega^2 - \Omega_\qq^2 - 2 \Omega_\qq g_\qq^2 \Pi(\qq,\omega)},
\end{align}
and leads to a softening of the phonon at that momentum
as shown in Fig.~\ref{fig:kohn_D} where we plot the phonon
spectral function.  As the electron-phonon coupling is increased,
a sharp dip develops in the phonon frequency, which eventually
nears zero. This is a mechanism for the formation of a charge density wave (CDW), but
in this section we will remain in the disordered phase.

In turn, the electron-electron interaction is
\begin{align}
    V_\qq(\omega) = |g_\qq|^2 D(\qq,\omega).
\end{align}
We evaluate the polarizability $\Pi(\qq,\omega)$ for the electrons using the Lindhard
formula\cite{mihaila2011lindhard}, i.e. the bare bubble.

\end{document}